\documentclass[usenatbib,usegraphicx,useAMS]{mn2e}
\usepackage{amsmath}
\usepackage{amssymb}
\usepackage{amsfonts}

\title[LGRBs and their host galaxies]{The host galaxies of long-duration GRBs in a cosmological
 hierarchical scenario}
\author[Nuza et al.]{
\parbox[t]{\textwidth}{
Sebasti\'an E. Nuza$^{1}$\thanks{E-mail: sebasn@iafe.uba.ar},
Patricia B. Tissera$^{1}$,\thanks{E-mail: patricia@iafe.uba.ar},
Leonardo J. Pellizza$^{1,2}$, 
Diego G. Lambas$^{3}$, 
Cecilia Scannapieco$^{1}$\thanks{Current address: Max-Planck-Institut f\"ur Astrophysik, Karl-Schwarzschild-Str. 1, D-85748 Garching, Germany} 
and Mar\'{\i}a E. De Rossi$^{1}$
}
\vspace*{6pt} \\ 
$^{1}$Instituto de Astronom\'{\i}a y F\'{\i}sica del Espacio, C.C. 67, Suc. 28, 1428, Buenos Aires, Argentina\\
$^{2}$AIM (UMR 7158), Service d'Astrophysique, CEA Saclay, B\^at. 709, L'Orme des Merisiers, 91191 Gif-sur-Yvette, France\\
$^{3}$Grupo IATE, Observatorio Astron\'omico de C\'ordoba, Laprida 854, X5000BGR, C\'ordoba, Argentina
\vspace{-0.5cm} 
}
\begin{document}

\date{Accepted  Received ; in original form }


\maketitle

\label{firstpage}

\begin{abstract}
We developed a Monte Carlo code to generate long-duration gamma ray burst (LGRB) events within cosmological hydrodynamical simulations consistent with the concordance $\Lambda$CDM model. As structure is assembled, LGRBs are generated in the substructure that formed galaxies today. We adopted the collapsar model so that LGRBs are produced by single, massive stars at the final stage of their evolution. We found that the observed properties of the LGRB host galaxies (HGs) are reproduced if LGRBs are also required to be generated  by low metallicity stars.
The low metallicity condition imposed on the progenitor stars of  LGRBs selects a sample of HGs with mean gas 
  abundances of 12 + log O/H $\approx 8.6$.
For $z<1$ the simulated HGs of low metallicity LGRB progenitors tend to be faint, slow rotators with high star formation efficiency,
compared with the general galaxy population, in agreement 
  with observations. At higher redshift, our results suggest  that larger systems with high star formation activity could also contribute to the generation of LGRBs from low metallicity progenitors since the fraction of  low metallicity gas available for star formation increases for all systems with look-back time. 
Under the hypothesis of our LGRB model, our results support the claim that LGRBs could
be unbiased tracers of star formation  at high redshifts.
\end{abstract}


\section{Introduction}

Gamma ray bursts (GRBs) are the most energetic electromagnetic events in the Universe (e.g. Piran 2000; M\'esz\'aros 2002). Different models have been developed to explain the physical origin of these bursts. It is now widely believed  that  LGRBs are principally associated with the star core collapse into a black hole during a Type Ib/c Supernova (SN) event (MacFayden, Woosley \& Heger 2001). 
This model, also known as the collapsar scenario, is linked to the evolution of single, massive stars which  have  mean lifetimes of several million years, implying a negligible typical lifetime for LGRB progenitors in cosmological terms. 
From this point of view, these events are commonly considered as possible unbiased tracers of the cosmic star formation history up to high redshift. However, this is still a controversial point due to different sources of bias such as
dust extinction intrinsic to the host (e.g. Jakobsson et al. 2005; Priddey et al. 2006). 
Also, the presence of metals in the progenitor star could play a significant role in the generation of LGRBs, introducing a bias. The metallicity of the progenitor star is an  important input parameter for the collapsar model. High metallicity stars are able to develop strong stellar winds which lead to the loss of both mass and angular momentum. In the case of the collapsar model, this can prevent the generation of LGRBs (e.g. MacFadyen \& Woosley 1999).
In this model, GRBs can only be formed by massive, single stars with a metallicity below $ 0.3 Z_{\odot}$
(Hirschi et al. 2005; Woosley \& Hegger 2006).

If GRBs are associated to the formation of massive stars, they might  provide information on the star forming regions in galaxies at different redshifts (e.g. Mao \& Mo 1998; Hogg \& Fruchter 1999). However, only recently it has been possible to obtain more insight into the properties of the HGs as discussed by Le Floc'h et al. (2003, hereafter LF03) and  Savaglio, Glazebrook \& Le Borgne (2006), among others. 
These works find a trend for the HGs to be sub-luminous and bluer with higher star formation activity than luminous
infrared  galaxies. 
Recent observations also provide information on the chemical abundances of the HGs at different redshifts (Sollerman et al. 2005;
Savaglio et al. 2006; Stanek et al. 2006; Wolf \& Podsiadlowski 2006; Fynbo et al. 2006). 
These results  suggest
that HGs could be low metallicity galaxies compared with star-burst galaxies  (see also Fruchter et al. 2006).
Conselice et al. (2005) studied a sample of 37 HGs of LGRBs up to $z\approx 3$, finding that, at $ z<1$,
 HGs seem to be smaller than the average galaxy population while, at higher redshifts, 
HGs are more similar to the general galaxy population.  

The recent detection of  LGRBs at $z \approx 6$ probes that these events could also provide a unique physical technique for the identification of galaxies at very early stages of their evolution, which can not be spotted with usual methods (Tagliaferri et al. 2005; Berger et al. 2006).
The combination of  this information with other observational results for high redshift galaxies (e.g. Chen et al. 2005; Starling et al. 2005) opens the possibility for a new stage in the study of the chemical enrichment of the Universe by extending our knowledge to a very early epoch of galaxy formation.

Galaxy formation models could help us to test different possible GRB scenarios, provided that the connection between the star formation activity  and the triggering of GRB events is valid.
Within a cosmological framework (in particular in the $\Lambda$CDM concordance model), the formation of galaxies is a complex process where the structure is assembled in a highly non-linear way. 
Numerical simulations have proven to be a powerful tool to study the formation of structure in this regime. In particular, smoothed particle hydrodynamical codes are able to describe the joint evolution of dark matter and baryons while allowing the treatment of specific processes such as star formation, chemical evolution  and SN energy feedback  (e.g. Katz \& Gunn 1991; Navarro \& White 1993; Mosconi et al. 2001; Springel \& Hernquist 2003; Scannapieco et al. 2005, 2006).  
Recently, Courty, Bj\"ornsson \& Gudmunsson (2004) made use of hydrodynamical structure formation simulations with the aim at identifying galaxy populations capable of reproducing the general observational features of the HGs. These authors find that, by requiring the HGs to have high star formation efficiency, the observed HG luminosity distribution could be reproduced.

In this work, we developed a Monte Carlo code for simulating the triggering of LGRB events 
 assuming the collapsar model for their progenitor stars as described by Fryer, Woosley \& Hartmann (1999, hereafter FWH99). 
The LGRB code is then coupled to hydrodynamical cosmological simulations so that the generation of LGRBs can be followed as
 the structure forms and evolves. We used hydrodynamical simulations performed with the chemical GADGET-2 code of Scannapieco et al. 
(2005). This code can follow the chemical enrichment of baryons as the structure forms. The dynamical and chemical properties of the 
simulated HGs are analyzed and confronted with available observations. However, the simulations analyzed in this paper do not include 
SN energy feedback. This process is expected to be able to trigger strong, metal-loaded, galactic winds capable of transporting material 
out of the galaxies, affecting the dynamics of the systems. Although SN energy feedback is an important physical mechanism, 
its modellization in hydrodynamical simulations is still controversial and only recently new numerical algorithms seem to be able to 
self-consistently take it into account (Marri \& White 2003; Scannapieco et al. 2006). For this reason, in this work, we concentrate on
 the chemical enrichment of baryons which is regulated by the gravitational growth of the structure and gas cooling, leaving for a future 
work the treatment of SN energy feedback. Considering the critical dependence of the collapsar model on the metallicity of the progenitor 
star, the possibility of estimating the metallicity of the simulated stars self-consistently provides us with a powerful tool to study LGRB host galaxies.

The present paper is organized as follows: Section 2 describes the simulated galaxy catalogues used in this work. Section 3 presents the  Monte Carlo code for simulating the LGRBs. Section 4 provides the analysis and discussion whereas Section 5 summarizes the main results.

\begin{figure}
\begin{center}
\includegraphics[width=8.5cm]{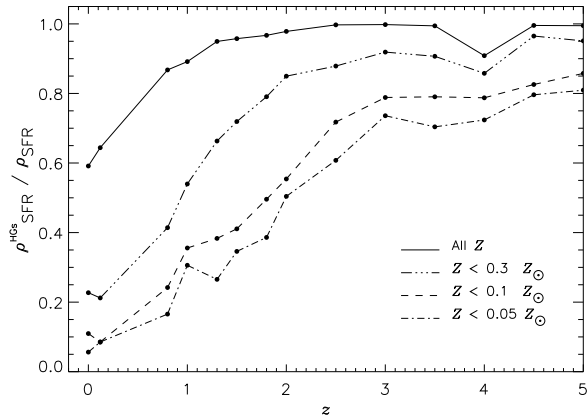}
\caption{Ratio between the cosmic star formation rate density estimated from the the simulated HGs and the total one of the simulated box ($\rho_{\rm SFR}$) as a function of redshift. For Model II, we include three different predictions by using different metallicity thresholds for the progenitor stars.}
\label{fig1}
\end{center}
\end{figure}

\section{The simulated galaxy catalogue}

We used the synthetic galaxy catalogues constructed by  De Rossi, Tissera \& Scannapieco (2006) from a hydrodynamical cosmological simulation of a typical region of the Universe consistent with a concordance cosmogony with $\Omega_{\rm M}=0.3$, $\Omega_{\Lambda}=0.7$, $\Omega_{\rm b}=0.04$, $\sigma_8 = 0.9$ and $H_0 = 100h$ km s$^{-1}$ Mpc$^{-1}$ with $h=0.7$.
The simulations were run by  using the chemical {GADGET-2} code of Scannapieco et al. (2005), grafted into the fully conservative {GADGET-2} (Springel \& Hernquist 2002; Springel 2005). The chemical code describes the  enrichment of the interstellar medium (ISM) by Type II and Type Ia Supernovae (SNII and SNIa, respectively), assuming a standard Salpeter initial mass function (Salpeter 1955), with a lower and upper mass cut-offs of $0.1$ and 40 M$_{\odot}$, respectively. This code provides the history of enrichment of each element of fluid by individual chemical elements such as, e.g. Fe$^{56}$ and O$^{16}$ as the structure is assembled in consistency with the adopted cosmology. The interested reader is referred to Scannapieco et al. (2005; see also Mosconi et al. 2001) for more details on the chemical algorithms.

The simulated volume corresponds to a periodic box of 10 Mpc $h^{-1}$ 
comoving size resolved with  $2 \times 160^3$ total particles with an initial mass of $2.7 \times 10^6~{\rm M_{\odot}}~h^{-1}$ and 
$1.4 \times 10^7~{\rm M_{\odot}}~h^{-1}$ for the gas and dark matter particles, respectively
(simulation S160). We also analyzed a similar simulation (S80) with  $2 \times 80^3$ particles to test the effects of resolution ($1.8 \times 10^7~{\rm M_{\odot}}~h^{-1}$ and $1.4 \times 10^8~{\rm M_{\odot}}~h^{-1}$ for the gas and dark matter particles, respectively). For the S80 experiment, we found similar results to those of S160 supporting the claim that our findings are not strongly affected by numerical resolution (see Nuza et al. 2005 for details on the S80).
The size of the simulated  volume has been chosen as a compromise between 
the need to have a well-represented galaxy sample and
high enough numerical resolution to study the astrophysical properties of the
simulated galaxies. Certainly, we are representing only galaxies in the field. A larger
volume would be needed to study cluster regions which might contribute more importantly to GRBs  at high
redshift. However, we would not expect significant changes in our results since the largest differences
between the HGs and the general galaxy population are found for $ z < 1$, where most of the star formation activity
is located outside high density regions (e.g. Martinez \& Muriel 2006). 
However, we note that the small simulated volume makes the comparison with observations more complex since
as we move to high redshift, simulated galaxies get smaller as expected in a hierarchical clustering scenario.
Hence, at high redshift, the simulated galaxy population  differs from the observed samples which
tend to select luminous systems (e.g. Christensen et al. 2004; Fruchter et al. 2006). 

De Rossi et al. (2006) constructed the catalogues of simulated galaxies by  applying a density contrast criterion ($\delta \rho / \rho \approx 178~{\Omega_{\rm M}}^{-0.6}$ as in White, Efstathiou \& Frenk 1993) to identify virialized structures.
These catalogues include systems with virial mass larger than $2 \times 10^8~{\rm M_{\odot}}~h^{-1}$ in order to prevent strong numerical artifacts. The synthetic catalogues comprises galaxies from $z=3$ to $z=0$. Our numerical resolution prevents us from exploring higher redshifts.
For each simulated galaxy, the dynamical, astrophysical and chemical properties are estimated at the optical radius. This radius is defined as the one that encloses $83\%$ of the baryonic mass of the systems and provides a more realistic way of confronting simulated galaxies with observed ones. De Rossi et al. (2006) also estimated the synthetic magnitudes for each simulated galaxy by combining the age and metallicity of each star particle in the simulated galaxies and applying the population synthesis model of Bruzual \& Charlot (2006).

Hence, we will use the dynamical, astrophysical and chemical properties such as  optical  velocity $V_{\rm opt}$\footnote{The optical  velocity is the circular velocity (i.e. $\sqrt{GM/R}$) measured at the optical radius. This velocity is comparable to the rotational velocity measured for galaxies at the maximum observed radius.}, absolute magnitudes and mean O/H abundances for the simulated galaxy sample comprised by 227, 241, 202 and 137 objects for $z=0,1,2,3$ respectively with the purpose of studying the properties of the HGs.

\section{The long-duration gamma-ray burst monte carlo code}

A Monte Carlo code has been designed to work incombination with numerical simulations of a typical region of the Universe to generate LGRB events in each of the galaxies in the simulation. We consider a simulated galaxy with at least one LGRB event as HG.
We assumed Poissonian statistics to emulate the probability distribution function of these events. FWH99  studied several alternatives for the GRB  progenitors, concluding that the collapsar model was the most likely scenario (i.e. its expected rate is typically several orders of magnitude greater than for the coalescence-type GRB models).
For this scenario, FWH99 quoted an event rate of $R_{\rm coll} \sim 10-1000$ Myr$^{-1}$ Galaxy$^{-1}$, where Galaxy represents a typical galaxy.
In the collapsar model, the progenitor stars  are assumed to be massive stars, ending their lives as SNIb/c.
 Hence, our LGRB algorithm is designed to select massive stars as GRBs candidates which have a short life time.
 For this purpose, we checked all star particles within two optical radius of 
 a given simulated galaxy,  searching for
those with a stellar age younger than an adopted stellar-age cut-off, $t_{\rm c}$. 
The Salpeter initial mass function (see Section 2) provides
the number of massive stars within a given star particle. 
Therefore, only massive stars younger than $t_{\rm c}$ are considered as LGRB candidates.

The Monte Carlo model was tested with several reasonable $t_{\rm c}$ values within the limits imposed by stellar evolution, being the results insensitive to this parameter. Throughout this work, we assumed $t_{\rm c} = 5 \times 10^6$ yr, $R_{\rm coll}=100$ Myr$^{-1}$ Galaxy$^{-1}$ and a typical galaxy mass of $10^{11}~{\rm M_{\odot}}$. This set of parameters defines  Model I.

Based on the fact
 that the high metallicity of the progenitor envelope may prevent the triggering of GRBs within the collapsar model 
(e.g. Woosley \& Hegger 2006), we also developed a Model II which takes into account the metallicity of the progenitor stars. 
In fact, in the collapsar model,  $ Z =0.3 Z_{\odot}$ is  the maximum allowed metallicity for the triggering of GRBs.
 Hence, in Model II, only low metallicity, massive stars are considered as possible LGRBs progenitors.
We tested the sensitivity of the results to the adopted metallicity threshold and chose one as the reference value as discussed in next Section.

We applied these LGRB models to all simulated galaxies  from $z=3$ to $z=0$. We generated 1000 Monte Carlo realizations for every selected
 stellar population (represented by a given star particle) in each simulated galaxy of the catalogues for Model I and Model II, in order to 
generate probability distributions for LGRB events.

\section{Analysis and Results}

By applying our LGRB scheme to each of the simulated galaxies as a function of redshift, we generated a distribution of possible LGRB events. Since the LGRB scheme assumes that the progenitors are massive stars, then we expect that the simulated host galaxies of LGRB will reproduce the star formation rate (SFR) history of the simulated volume. And this is actually the case for Model I where the predicted HGs agree with the general galaxy population for $z >1$.
For $z <1$, we found a difference between the $\rho_{\rm SFR}$ determined by the total galaxy population and that traced by HGs in Model I. Up to $40\%$ of new stars are born in galaxies without LGRBs. This is a result of statistical fluctuations introduced by the low rates associated to LGRBs given the general decrease of the SF activity in galaxies for $z < 1$. We found that galaxies at $z=0$ without LGRB events have a SFR distribution with the mean at
$\langle SFR \rangle \approx 0.1 {\rm ~M_{\odot}~{yr}^{-1}}$ while HGs have a broader distribution with the mean at $\langle SFR \rangle \approx 0.7  {\rm ~M_{\odot}~{yr}^{-1}}$.
We acknowledge that part of these effects could be due to the fact that we are not considering SN energy feedback.
As a consequence, small systems tend to exhaust their gas reservoir too quickly as soon as they collapse, so that their SF activity, later on, is lower than expected. However, this effect cannot explain the trend by itself since galaxies of different virial masses contribute to both sorts of galaxy populations (i.e. with and without LGRBs).

 For Model II, the behavior is different. In  Fig.~\ref{fig1} we show the ratio between the comoving star formation rate density ($\rho_{\rm SFR}$) estimated by considering only the selected HGs in Model II and the total $\rho_{\rm SFR}$ obtained from the complete galaxy population. To illustrate the sensitivity of Model II to the adopted metallicity threshold, we included the predicted  $\rho_{\rm SFR}$ for Model II using three metallicity thresholds, corresponding to $Z=0.3Z_{\odot}$, $Z=0.1Z_{\odot}$ and  $Z=0.05Z_{\odot}$, respectively. 

Evidently, as the restriction in the metallicity increases, the difference between the predicted  $\rho_{\rm SFR}$ and the total $\rho_{\rm SFR}$ also increases. While for a threshold of $Z=0.3Z_{\odot}$, our simulations are capable of measuring $40-60\%$ of the star formation activity in the simulated volume at $z <1$, this percentage decreases to less than $20-40\%$ for a lower metallicity threshold. 
If the metallicity content of the stars is a key ingredient in the triggering of LGRBs, then tracing the star formation history of the universe by using these events may introduce a biased signal, principally for $z < 1$. However, at very high redshift, our results suggest that, at the utmost, $\approx 20 \%$ of the star formation activity could be missing if LGRBs are used as tracers of the star formation activity.
Hereafter, we will assume $Z=0.1Z_{\odot}$ as the reference value for the discussion.
Then Model II requires the  progenitor star to have a mean abundance 12 + log O/H $\leq 7.5$. This metallicity threshold corresponds to an O/H level approximately an order of magnitude lower than the solar one\footnote{ We assume
the solar abundance  12 + log O/H $\approx 8.7$ given by  Allende-Prieto, Lambert \& Asplund (2001).} and is consistent with the theoretical estimations of Yoon, Langer \& Norman (2006).

\begin{figure}
\begin{center}
\includegraphics[width=8.5cm]{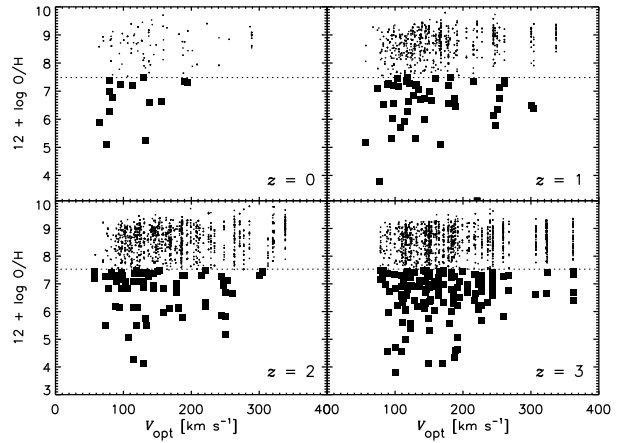}
\end{center}
\caption{Distribution of 12 + log O/H of  stars selected to produce LGRB events separated   in high (points) and low (filled squares) 
 metallicity LGRB  progenitors as a function of the optical velocity of their HGs at $z=0,1,2,3$.}
\label{fig2}
\end{figure}

\begin{figure}
\begin{center}
\includegraphics[width=8.5cm]{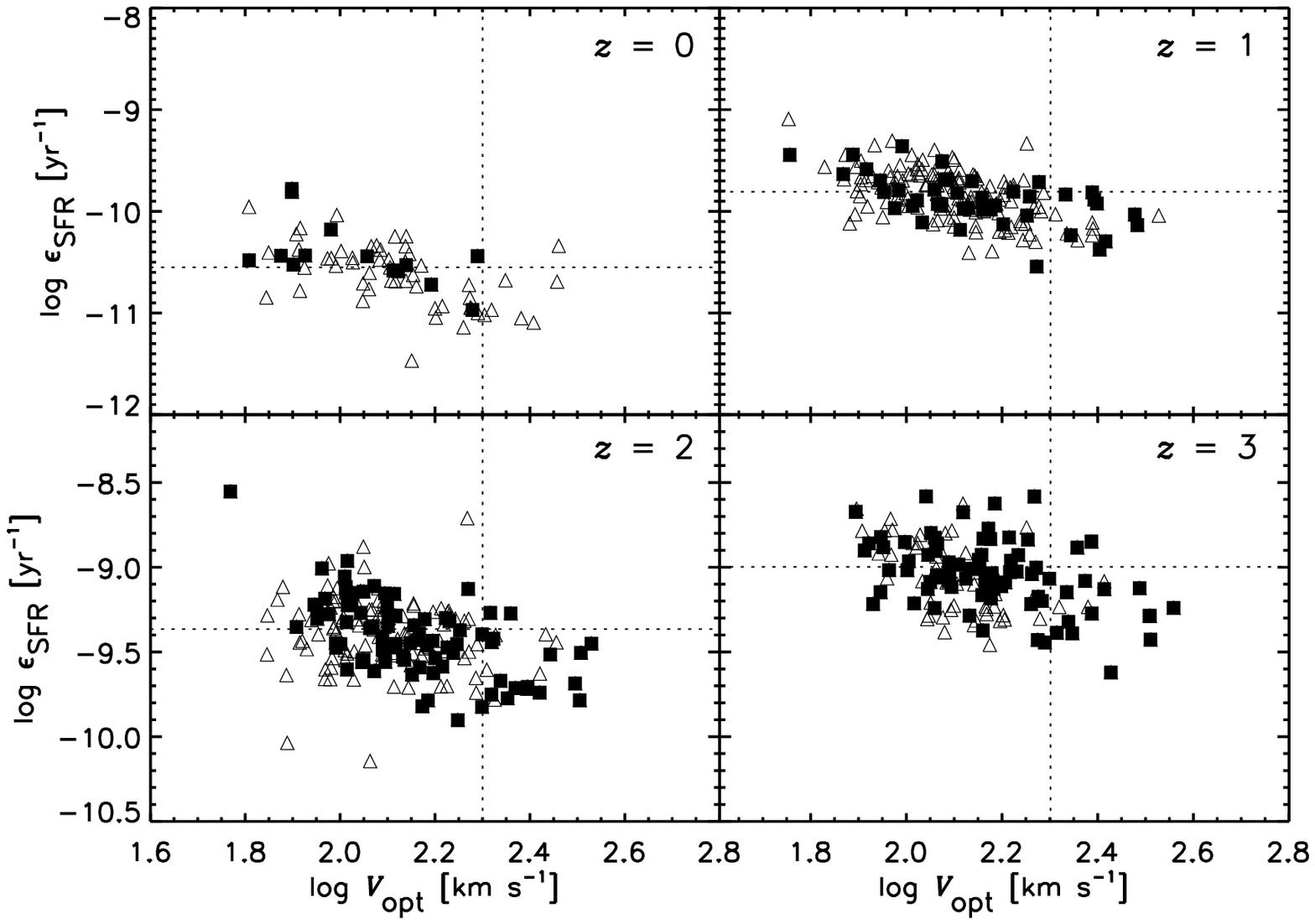}\\
\includegraphics[width=8.5cm]{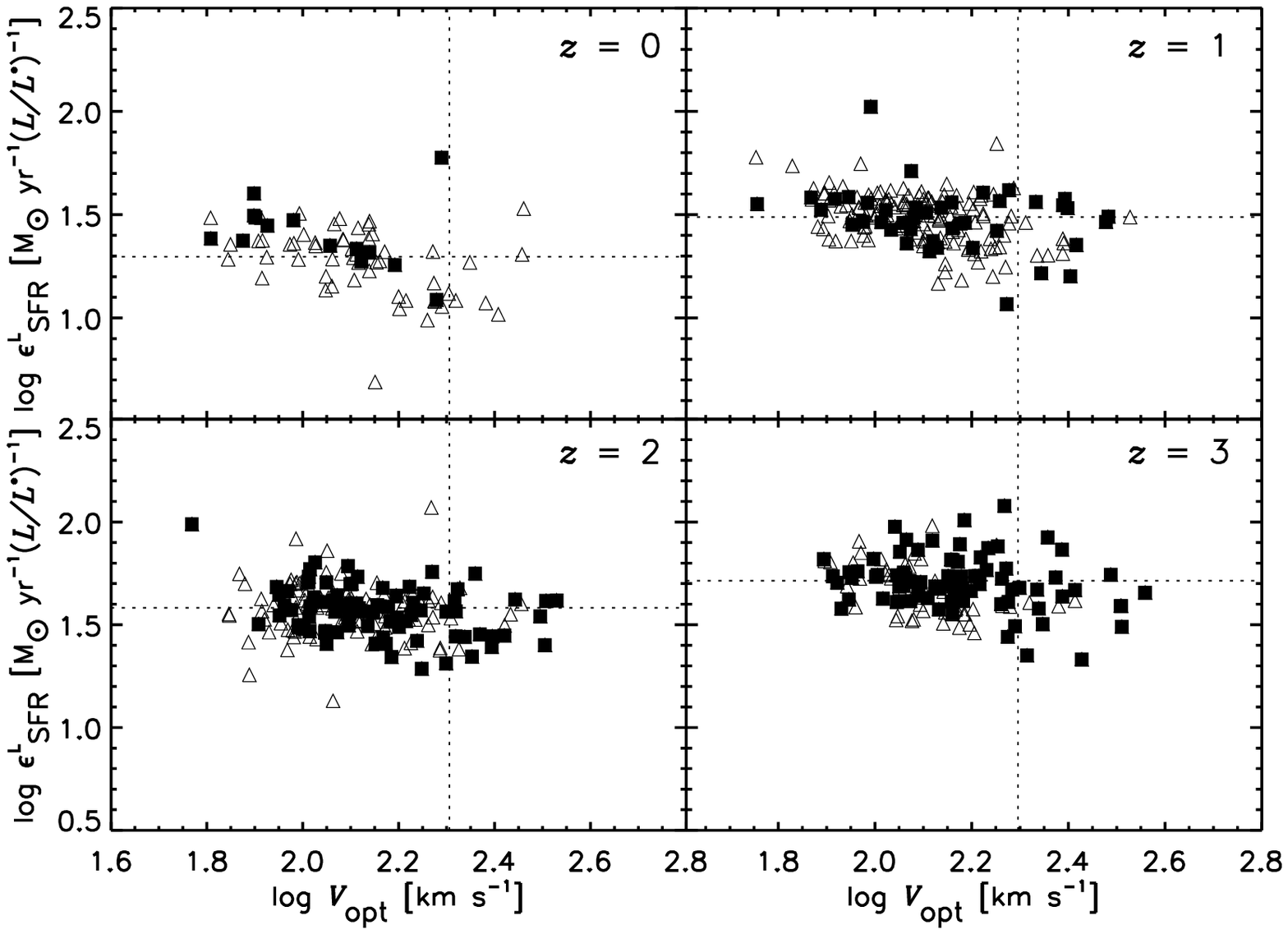}\\
\end{center}
\caption{Mean star formation rate efficiency $\epsilon_{\rm SFR}$ (top panel) and specific star formation rate $\epsilon^L_{\rm SFR}$ for the $B$-band (bottom panel) as a function of optical velocity $V_{\rm opt}$  for all the simulated HG sample in Model I (open triangles) at $z=0,1,2,3$. We also include the corresponding relation for HGs with low metallicity LGRBs progenitors of Model II (filled squares). Vertical dotted lines denote an optical velocity of 200 km s$^{-1}$ and horizontal dotted lines represents mean values of the general galaxy population.}
\label{fig3}
\end{figure}

\begin{figure}
\includegraphics[width=8.5cm]{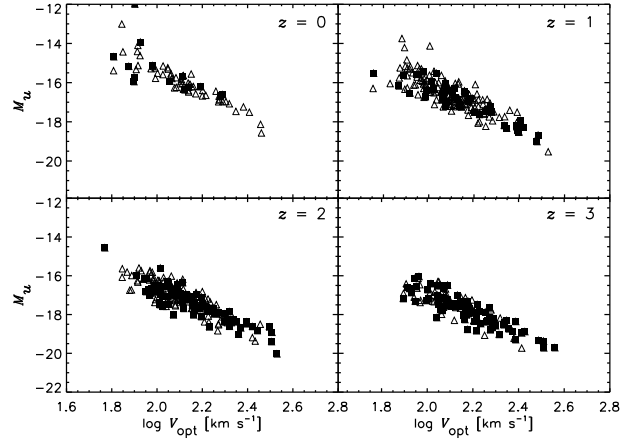}
\caption{u-band absolute magnitude as a function of optical velocities for the  HG samples in Model I (open triangles) and  in Model II (filled squares) at $z=0,1,2,3$. For clarity, we have not superimposed the general galaxy population which follows the same trends.} 
\label{fig4}
\end{figure}

\subsection{The simulated host galaxies}

In this Section, we  analyze the properties of the simulated HGs and compare them with available observations to investigate the nature of the systems which are more prompted to generate LGRBs as a function of redshift and within the hypothesis adopted in our models.
In Fig.~\ref{fig2} we show the distribution of O/H abundances of the LGRB progenitor stars as a function of $V_{\rm opt}$ of their HGs, distinguishing high-metallicity progenitors (points) from the low metallicity ones (filled squares).
The adopted metallicity threshold  is 12 + log O/H = 7.5 as discussed above. From this figure, it is clear that LGRBs tend to originate in HGs with $V_{\rm opt} \le  200~{\rm km~s^{-1}}$. Even more, we can  see that  Model II selects a subsample of LGRBs mainly originated in slow rotating systems. This is true for all the analyzed redshifts. The fraction of HGs with  $V_{\rm opt} > 200~{\rm km~s^{-1}}$ and with LGRB events originated in low metallicity stars are null,  0.16,  0.30,  0.24 for $z=0,1,2,3$, respectively (hereafter, we take this  set of redshifts as the reference one). The growing contribution of faster rotating systems with increasing redshift denotes a change in the composition of HGs.
 Note that, as shown in Fig.~\ref{fig1}, varying the metallicity threshold of Model II up (or down)   makes the HG samples 
 have more (or less) contributions from larger systems.

We also investigated a possible correlation between the mean O/H abundance of the HGs and those of LGRB progenitor stars.
We found a weak trend for the lowest metallicity progenitor stars to be located in the less enriched HGs.
In the case of Model II, the simulated HGs show even a weaker trend but these HGs have less mean abundances that
those of Model I. 
In Model II, the HGs show mean O/H abundances of 8.65, 8.67, 8.60 and 8.57 for the ISM and 8.64, 8.65, 8.57. 8.54 for the stellar populations at $z=0,1,2,3$, respectively. Within the hypothesis assumed in this paper, our results indicate that, the very low O/H abundance imposed on the progenitor star of a LGRB event does not imply  a very low mean metallicity content for their HGs. This is because the distribution of metals within
a given galaxy is not homogeneous but reflects the complex structure of the interstellar medium.  The mean  chemical
abundances of the simulated ISM of the HGs in Model II quoted 
above are consistent with the observed values reported by Sollerman et al. (2005), Savaglio et al. (2006), Stanek et al. (2006) and Wolf \& Podsiadlowski (2006) (see Section 4.3).

In Fig.~\ref{fig3} we show the SFR efficiency ($\epsilon_{\rm SFR}$) of the HGs defined as the ratio between the mean SFR and stellar
mass within the optical radius of simulated HGs  as a function of their optical velocity at $z=0,1,2,3$.
In this figure, we have over-plotted the values of $\epsilon_{\rm SFR}$ corresponding to Model I (open triangles) together with the relation estimated for HGs of low metallicity LGRBs (filled squares). As it can be seen, up to $z = 1$ the simulated HGs show a clear anti-correlation between $\epsilon_{\rm SFR}$ and $V_{\rm opt}$ so that lower rotating systems are more efficient at transforming their gas reservoir into stars than faster rotators. This trend is in agreement with observational results showing a similar behavior (e.g. Brinchmann \& Ellis 2000).
For $ z < 1$, we found that more than $50\%$ of the HGs in Model II have $\epsilon_{\rm SFR}$ larger than the average of the whole galaxy population. We note that this result agrees with the hypothesis made by Courty et al. (2004). 
From $z \approx 1$, the relation gets flatter, indicating roughly the same level of efficiency regardless of the galaxy circular velocity. Note that  the higher the redshift, the larger the global level of $\epsilon_{\rm SFR}$ since the simulated galaxies are more gas-rich with higher star formation activity as the redshift increases. It is worth mentioning again that the complete simulated sample of galaxies  have similar trends to that of Model I 
for all analyzed redshifts, principally for $z > 1$.

To better compare with  observations, we consider the total $B$-band luminosity instead of the stellar mass to estimate a SFR efficiency ($\epsilon_{\rm SFR}^L$), so-called specific SFR. We  used the luminosities estimated by De Rossi et al. (2006)  for the objects in this simulation. 
We  detected that HGs in Model II tend to have higher  
$\epsilon_{\rm SFR}^L$ than HGs in Model I and the general galaxy population, principally at low redshift. As discussed in Section 2,
the simulated galaxy population is made up of galaxies in the field, hence, at high redshift, we do not expect
very luminous and large systems to be included. This effect  makes the comparison with observations more complicated. 
Christensen et al. (2004) found that HGs are most likely similar to the field galaxies with the largest specific SFRs by carrying out an observational
study in the redshift range $0.43 \le z \le 2.04$. All their observed HGs have
$L/L^* \ge 0.07$ for the $B$-band (estimated using $M^{*}_B = -21$), while in our sample most of the simulated HGs have $L/L^* \lesssim 0.05$ and the general galaxy population has $75\%$ of the systems with $L/L^* \lesssim 0.07$.

The Tully-Fisher relation links the circular velocity of a galaxy with its luminosity (Tully \& Fisher 1977). In Fig.~\ref{fig4} we show the Tully-Fisher relation for the HGs in Model I and Model II at our four redshifts of reference. We used the luminosity in the  u-band of the Sloan Digital Sky Survey (hereafter SDSS).
 As it can be seen from this figure, the simulated galaxies determine a correlation at all analyzed redshifts, which indicates that  fainter systems are  slower rotators. Hence, considering the findings of Fig.~\ref{fig3} and Fig.~\ref{fig2}, we conclude that the HGs of LGRBs in Model II tend to be faint, slow rotating systems with  high star formation efficiency for $z < 1$
when compared to the general galaxy population. For higher redshift,
the predicted HGs in Model II cover the whole range of circular velocities and  high
star formation efficiencies. 
At $z >1$ we found no difference between the general galaxy population and HGs in Model I or Model II.

\begin{figure*}
\begin{center}
\includegraphics[width=8.5cm]{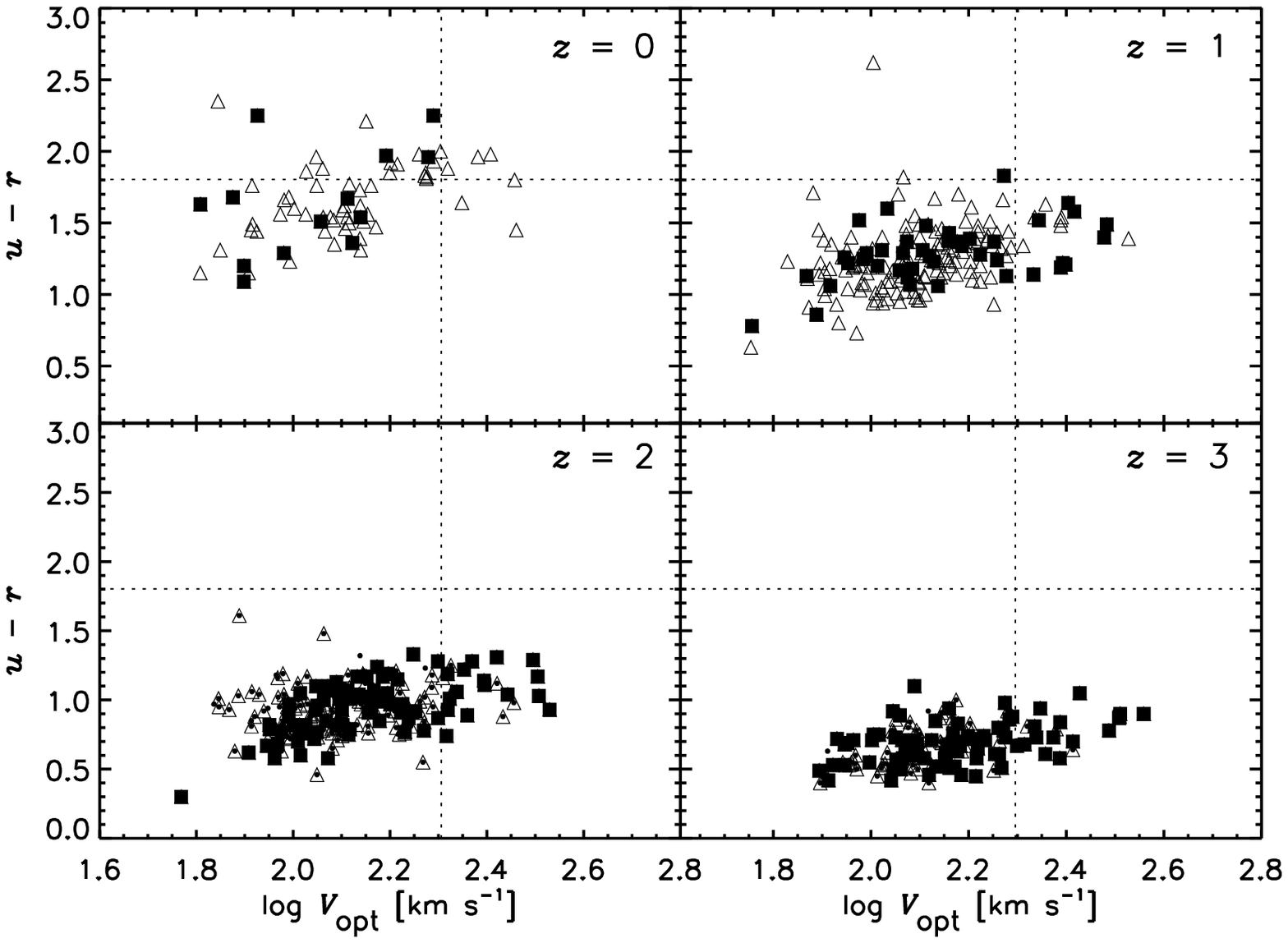}
\includegraphics[width=8.5cm]{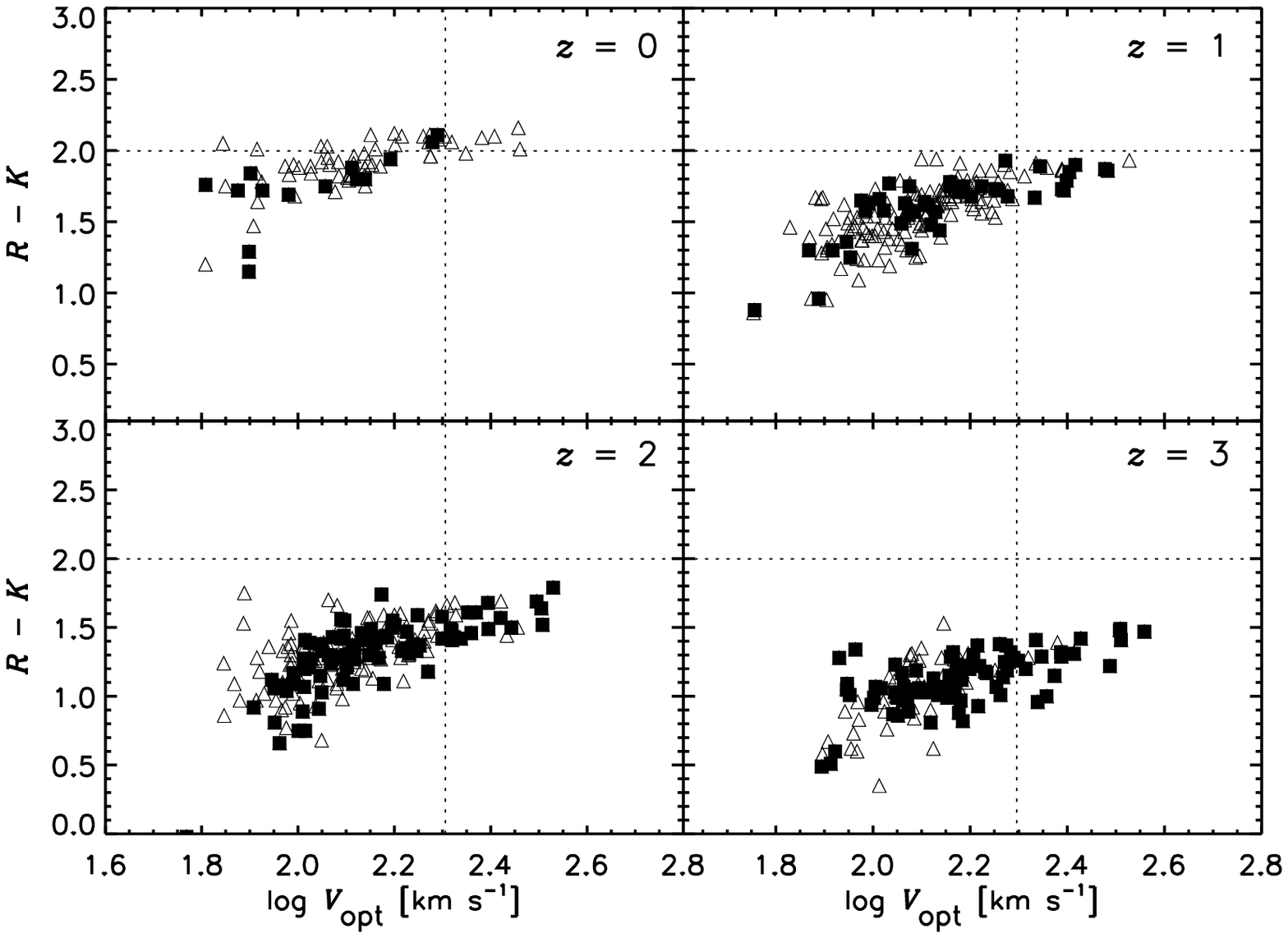}
\end{center}
\caption{$u-r$ (left panel) and $R-K$ (right panel) colours as a function of optical velocities of  HGs in Model I (open triangles) and  in Model II (filled squares) at the four  redshifts of reference. Vertical dotted lines denote an optical velocity of 200 km s$^{-1}$ and horizontal dotted lines represents $u - r = 1.8$ and $R - K = 2$ respectively.}
\label{fig5}
\end{figure*}

\subsection{Luminosity properties}

In this Section we analyze those HG properties that can be directly compared to HG observations. LF03 found that HGs tend to be blue and sub-luminous compared to luminous infrared galaxies.
We also analyzed the range of $u-r$ colours covered by the simulated HGs as shown in Fig.~\ref{fig5} (left panel). In the case of Model I, the distribution is similar to that defined by the complete simulated galaxy sample. However, HGs in Model II are restricted to
the blue end with most of the simulated HGs,  exhibiting $u-r < 1.8$ for  $z < 1$. We note that at high redshift  
the whole sample gets bluer since simulated galaxies are dominated by younger stars. Similar trends are 
found for  $R-K$ colours as shown in Fig.~\ref{fig5} (right panel).

\begin{figure}
\begin{center}
\includegraphics[width=8.5cm]{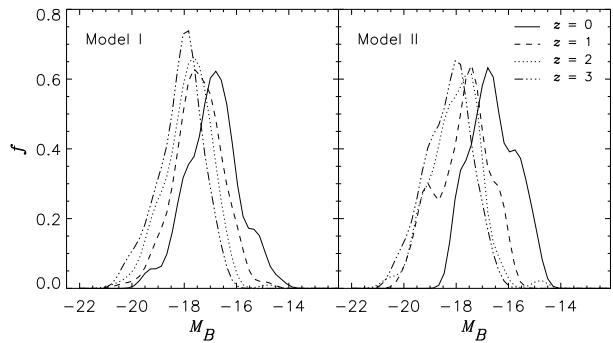}
\end{center}
\caption{Distributions of the absolute $B$ magnitudes for the sample of simulated HGs in Models I and II  at the four redshifts of reference. It can be seen that Model II exhibits a clear shift toward fainter luminosities for $z<1$.}
\label{fig6}
\end{figure}

\begin{figure}
\begin{center}
\includegraphics[width=8.5cm]{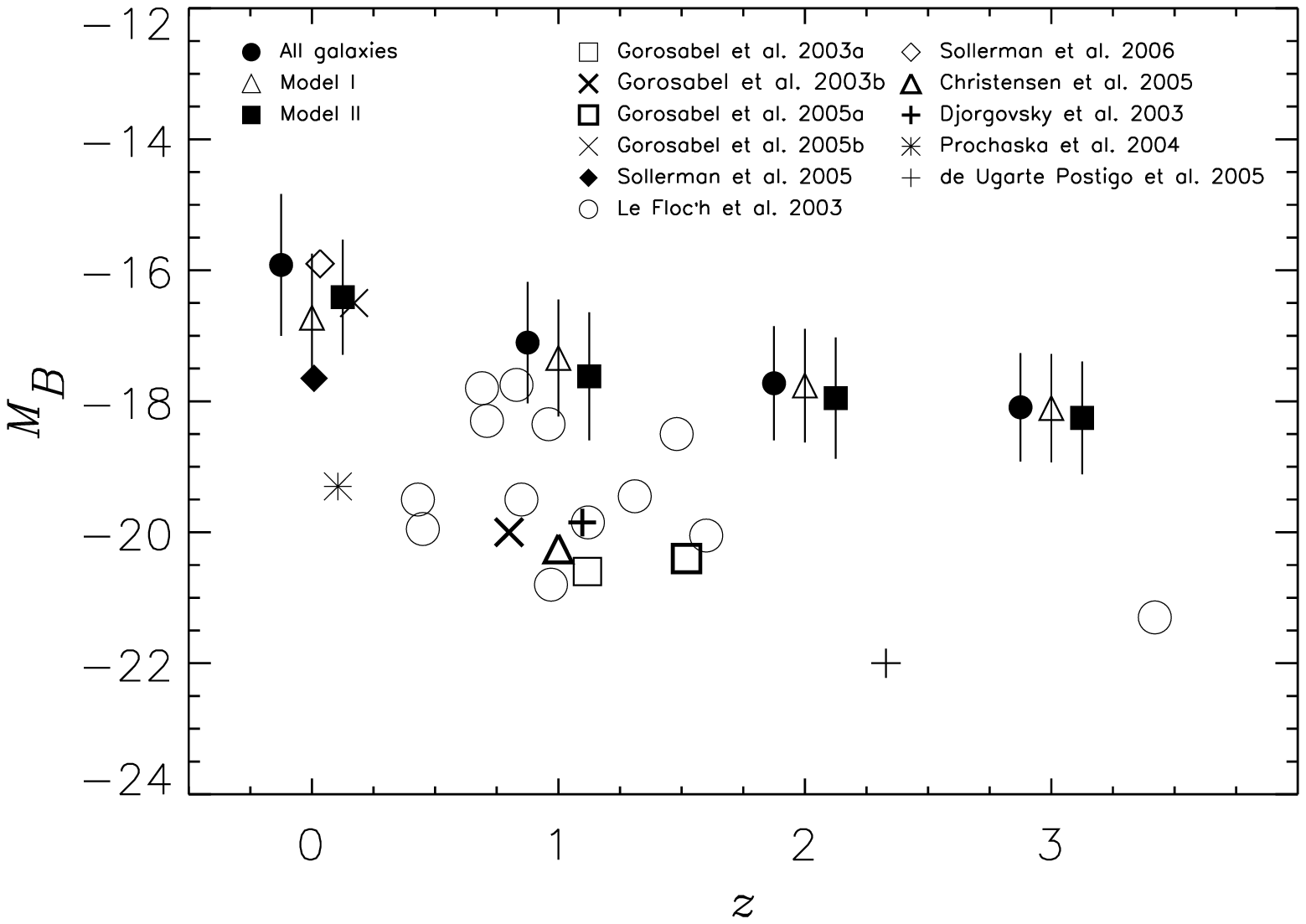}
\includegraphics[width=8.5cm]{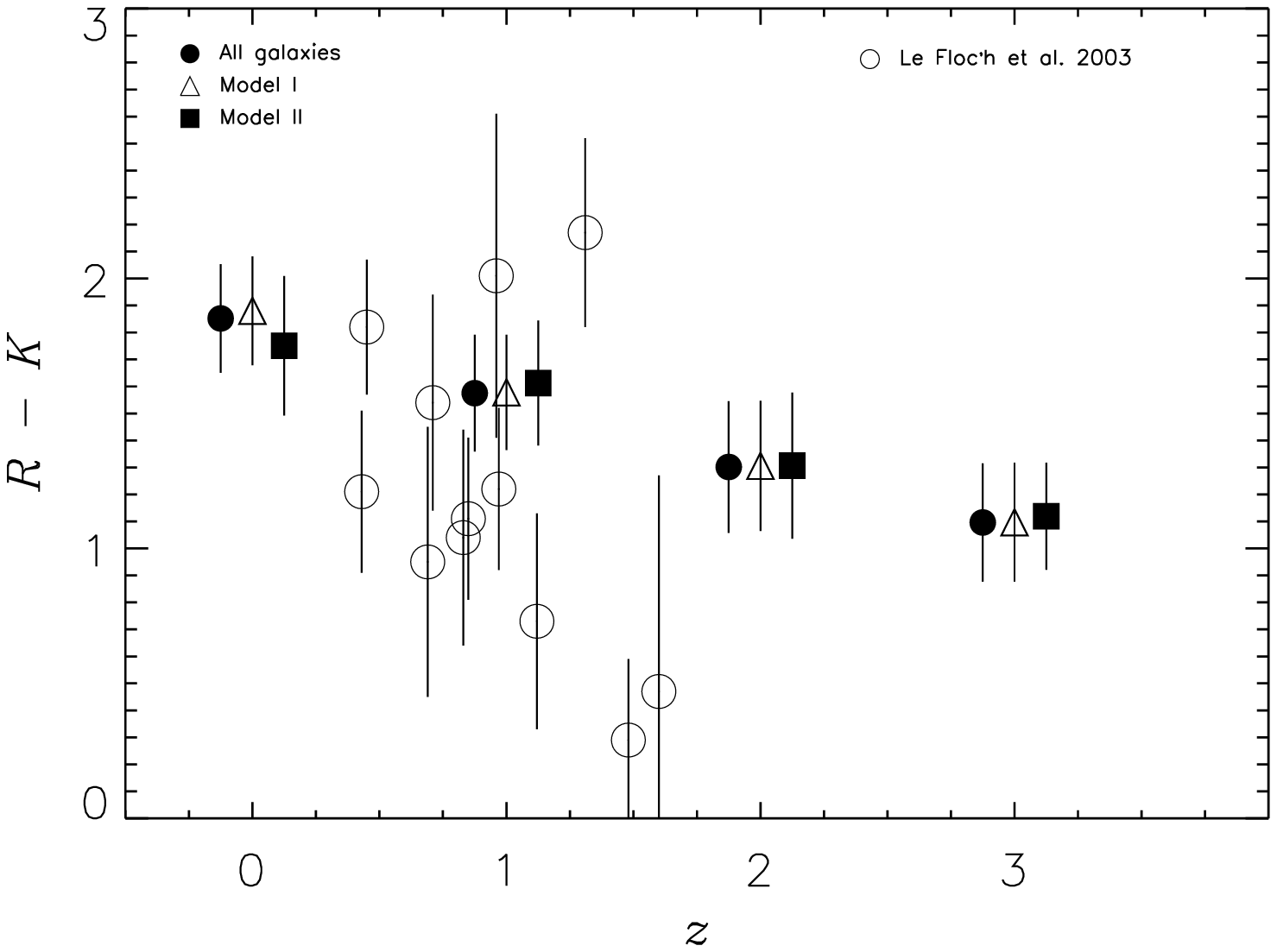}
\end{center}
\caption{Mean absolute $B$ magnitudes (upper panel) and the rest frame $R-K$ colours (bottom panel) for the simulated HG sample in Model I (triangles) and Model II (filled squares) as a function of redshift. For comparison we included different available observations of HGs. The error bars correspond to the standard deviation in the corresponding redshift interval. We have applied a small displacement in $z$ between the models in order to differentiate them.}
\label{fig7}
\end{figure}

 Fig.~\ref{fig6} shows the distributions of the  $B$-band absolute magnitude for the HG samples  in both models at the four redshifts of reference. As we can clearly see from this figure, the distributions shift toward brighter magnitudes  for higher redshift, in agreement with the results of LF03. This is a consequence of the fact that systems are dominated by younger stellar populations at higher redshift.
This trend is confirmed by Fig.~\ref{fig7} (upper panel) where the  $B$-band absolute magnitude is plotted as a function of redshift for several observed HGs (Djorgovsky et al. 2003; Gorosabel et al. 2003a,b; Prochaska et al. 2004; Christensen et al. 2005; Gorosabel et al. 2005a,b; de Ugarte Postigo 2005; Sollerman et al. 2005, 2006) also including those HGs of LF03 and the mean values for the simulated HGs in Model I and Model II.  Both models differ slightly in their mean values for $z > 0$, showing the main difference (roughly 1 mag) at $z <1 $. As it can be seen in the figure the observed HGs are within $\sim 2\sigma$ of the simulated HGs. However, we note that the simulated hosts are systematically fainter than the observed ones. This could be due to the combination of two effects. First, we are simulating a small volume of a typical field region where very bright
galaxies are not expected to inhabit. Secondly, observations could be biased toward the detection of  brighter HGs with increasing redshift, while in the simulations, we get smaller systems with higher redshifts as a consequence of the hierarchical clustering. In the bottom panel of Fig.~\ref{fig7}, we show the rest frame $R-K$ colours for the objects in models I and II compared to the $k$-corrected LF03 sample. The colours of the simulated systems are within the observed range. Nevertheless, we note that they show less scatter and tend to be redder than observed HGs at a given redshift.

\begin{figure}
\begin{center}
\includegraphics[width=8.5cm]{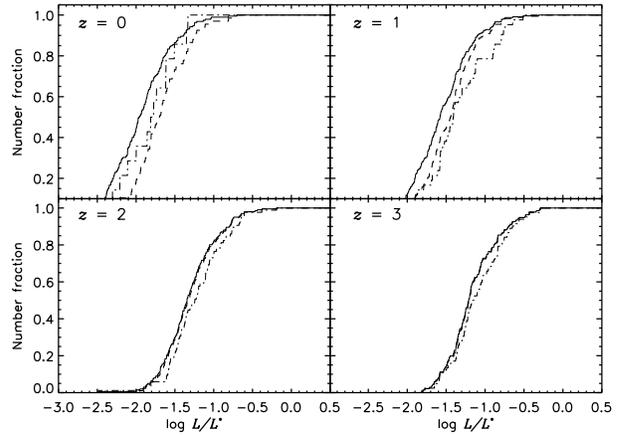}
\end{center}
\caption{Cumulative $B$ luminosity normalized to the characteristic luminosity of galaxies at $z=0$ for the simulated HGs in Model I (dashed line),  Model II (dotted-dashed line) and for the general galaxy population (solid lines) at $z=0,1,2,3$.}
\label{fig8}
\end{figure}

Another way of quantifying the differences between the total simulated HG sample (Model I) and the low metallicity one (Model II) is to estimate the fraction of cumulative luminosity  for both samples. In Fig.~\ref{fig8}, we  display these relations from $z=0$ to $z=3$  normalized to 
the characteristic luminosity $L^*$. 
It is clear that HGs in Model II contribute with fainter systems in comparison to those of Model I or to the general galaxy population, although for $z > 1$, the distributions are more similar. 
If luminosity is exchanged by stellar mass, we find that the low metallicity HG sample (Model II) contributes with smaller stellar mass systems than the complete HG sample (Model I) at low redshift. Hence, in these simulations, HGs of low metallicity LGRBs tend to  be smaller systems with high star formation efficiency compared to the global galaxy population at $ z < 1$.

\subsection{The stellar mass-metallicity relation}

\begin{figure}
\begin{center}
\includegraphics[width=8.5cm]{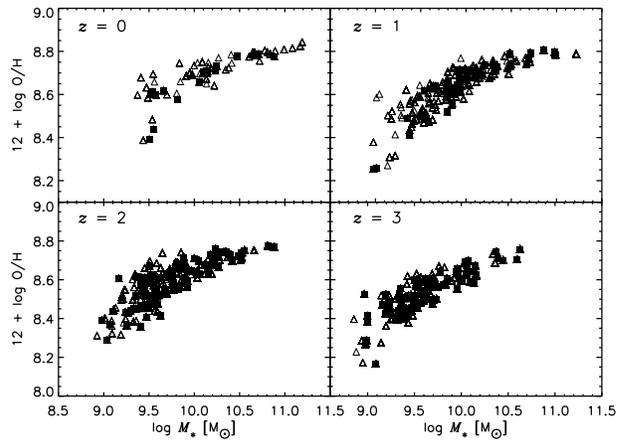}
\end{center}
\caption{Stellar mass-metallicity relation for HGs in Model I (open triangles) and Model II (filled squares) at $z=0,1,2,3$. For clarity, we have not superimposed the general galaxy population which follows the same trends.}
\label{fig9}
\end{figure}

The stellar mass-metallicity relation (MZR) is a well-defined observed relation which has been determined
at low (e.g. Tremonti et al. 2004) and high redshift (e.g. Erb et al. 2006). 
 We estimated the simulated MZR for the HGs in Model I and Model II as a function of redshift. 
We found that both samples follow the general MZR predicted for the simulated galaxy population (Tissera, De Rossi \& Scannapieco 2005). 
As it can be seen from Fig.~\ref{fig9}, at $z=0$, only 23 $\%$ of the HG sample in Model II has a stellar mass larger than the characteristic mass of $10^{10.2}~{\rm M_{\odot}}~h^{-1}$. This characteristic mass was  determined by Tissera, De Rossi \& Scannapieco (2005) and is in rough agreement to that estimated observationally by Kauffmann et al. (2004) by studying galaxies in the SDSS. 
 Tremonti et al. (2004) found a  change in the slope of the observed MZR   at this characteristic value.
 As one moves to higher redshifts, more LGRBs with low metallicity progenitors are produced in larger stellar mass systems. In general, we can say that in our models the LGRBs pick galaxies that trace the general MZR, particularly at $z > 0$.

We note that if the metal content of the LGRB progenitor star is used as a tracer of the mean stellar metallicity of the systems instead of the mean metallicity of the HGs, the dispersion introduced is so large that the MZR is lost. At least in our models, the chemical abundance of the progenitor stars are not a good measure of the mean metallicity of the HGs.

\section{Conclusions}

We developed a Monte Carlo code in order to generate LGRBs adopting the event rate inferred for the collapsar model in FWH99. This code works in combination with hydrodynamical cosmological simulations  consistent with a $\Lambda$CDM scenario. We assumed Poissonian statistics for the probability distribution function of the events and  adopted a collapsar rate $R_{\rm coll}$ from FWH99. Two LGRB models have been developed. Model I requires the progenitor star to be a massive one. This model yields a HG population with similar properties to those of the total simulated galaxy sample. 
Model II includes an additional condition on the metallicity of
the progenitor star: only low metallicity, massive stars are taken as LGRB candidates, adopting a metallicity threshold an order of magnitude
lower than solar as a reference value.
 We found that Model II produces a set of HGs with properties in general agreement with observed ones. According to this model, HGs tend to be slow rotating systems with high star formation efficiencies,  principally at $z < 1$. At higher redshift, the correlations get weaker since all  galaxies tend to  have high star formation efficiencies and
larger  low-metallicity gaseous reservoirs. 
These star formation efficiencies have been estimated by using the stellar mass and  the $B$-band luminosity.

While  at all redshifts there is a trend to have LGRBs in faint, small stellar mass systems with circular velocity smaller than 200 km $\rm{s^{-1}}$, for $z>1$, larger systems can also significantly contribute. This is produced by the larger disponibility of low metallicity gas for star formation in systems of all masses at high redshift.
According to our results, at $z < 1$, HGs in Model II would tend to be smaller systems than the general galaxy population while
at higher redshift,  systems of different morphology and characteristics could be HGs. These results
are in agreement with recent observational findings discussed by  Conselice et al. (2005).

We found no tight correlation between the O/H abundance of the LGRB progenitor star and the mean  metallicity of the HGs. However, in Model II the simulated HGs tend to have mean abundances smaller than the 
solar one since most of the LGRBs are produced in the small, gas-rich systems in agreement with observations. 
The low metallicity condition for the progenitor star resulted to be a key ingredient in our model
to  match the observed properties of HGs at low redshifts ($z< 1$).  
In Model II, the mean metallicity of the HGs are also within the observed values lately reported, although we have some
simulated HGs with higher metallicities.  It is also important to note that Model II is able to reproduce the main characteristics
of the observed HG of LGRBs without getting  contradiction with  theoretical  models for LGRBs which require metallicities smaller  than $\sim 0.3$ solar. 

The estimation of the MZR  for HGs shows that in Model II,  these systems  are also good tracers of the galaxy MZR, although at $ z < 1$ this model  tends to map better the low mass end. In general we found that HGs do not differ significantly from
the general galaxy population although the probability to have a LGRB event is higher in galaxies with larger star formation
activity and in faint systems (since they are also more abundant and have more metal-poor gas).
In our simulations, the MZR is determined by the transformation of gas into stars  which is mainly regulated by the assemble of the structure (De Rossi et al. 2006) and its continuous enrichment by previous stellar generations. 
Since no SN winds have been modeled in this work, we can not  study their effect on the HGs relations.
 It is expected that SN winds might modify  the distribution of chemical elements  in galaxies, preferentially in low mass systems (e.g Larson 1974) where powerful, metal-loaded winds can  develop (Scannapieco et al. 2006).
However, we do not expect that this process would significantly modify our results,  since
slow rotating systems have, nevertheless, low metallicity. For this reason, we actually expect the predicted trends to be even stronger
when including SN energy feedback since this process will help to regulate the transformation of gas into stars and  will also eject part of the metals into the intergalactic media preferentially  for slow rotating systems. Although it will be interesting to study the effects of  SN energy feedback in our results in the future, it is encouraging that this simple scheme for LGBRs,  consistently included within a cosmological simulation, can already reproduce general trends of observed HGs.

\section*{Acknowledgments}
We thank the anonymous referee for useful comments which helped to improve this paper significantly.
Numerical simulations were run on Ingeld and HOPE PC-clusters at the Instituto de Astronom\'{\i}a y F\'{\i}sica del Espacio (IAFE). We acknowledge financial support from the Argentine institutions: CONICET, Fundaci\'on Antorchas, ANPCyT, SECyT-UNC and from the European Commission's ALFA-II programme through its funding of the Latin-american European Network for Astrophysics and Cosmology (LENAC).

\end{document}